\documentclass[prb,showpacs,preprintnumbers,amsmath,amssymb,twocolumn]{revtex4}
\usepackage{color}
\usepackage{graphicx}
\usepackage{epsfig}

\begin{document}
\title{Theory of the Raman spectra of the Shastry-Sutherland antiferromagnet SrCu$_2$(BO$_3$)$_2$ doped with nonmagnetic impurities }

\author{S. Capponi}
\email{capponi@irsamc.ups-tlse.fr} 
\affiliation{Universit\'e de Toulouse; UPS; Laboratoire de Physique Th\'eorique 
(IRSAMC);  F-31062 Toulouse, France}
\affiliation{CNRS; LPT (IRSAMC); F-31062 Toulouse, France}
\author{D. Poilblanc}
\affiliation{Universit\'e de Toulouse; UPS; Laboratoire de Physique Th\'eorique 
(IRSAMC);  F-31062 Toulouse, France}
\affiliation{CNRS; LPT (IRSAMC); F-31062 Toulouse, France}
\author{F. Mila}
\affiliation{Institute of Theoretical Physics, Ecole Polytechnique F\'ed\'erale de Lausanne,
1015 Lausanne, Switzerland}
\affiliation{}

\begin{abstract}
Controlled doping of SrCu$_2$(BO$_3$)$_2$, a faithful realization of the 
Heisenberg spin-1/2 antiferromagnet on the Shastry-Sutherland lattice, 
with non-magnetic impurities generates bound-states below the spin gap. 
These bound-states and their symmetry properties are investigated by exact 
diagonalisation of small clusters and within  a simple effective model describing a 
spinon submitted to an attractive extended potential.
It is shown that Raman spectroscopy is a unique technique to probe these bound-states. 
Quantitative theoretical Raman spectra are numerically obtained. 
\end{abstract}

\date{\today}

\maketitle

\section{Introduction and motivations}
SrCu$_2$(BO$_3$)$_2$ is an experimental realization of a spin 1/2 antiferromagnet 
living on the two-dimensional (2d) Shastry-Sutherland lattice (SSL). 
Its properties can be well described with the Heisenberg model:
\begin{equation}
{\cal H} = J \sum_{nn} {\bf S}_i \cdot {\bf S}_j +  J' \sum_{nnn} {\bf S}_i \cdot {\bf S}_j
\end{equation}
where $J$ (resp. $J'$) is the exchange within (resp. between)
dimers. The experimental compound has a ratio $\alpha=J'/J$ slightly
above~\cite{Miyahara2003} 0.6.  The SSL has been first introduced theoretically as an example of a 2d 
antiferromagnet whose ground-state for
small enough $\alpha$ is exactly known~\cite{Shastry1981}: It is simply the product of singlets on each dimer up 
to~\cite{Miyahara2003} $\alpha\simeq 0.7$. 
Indeed, experiments on SrCu$_2$(BO$_3$)$_2$ indicate a finite spin gap. For larger $\alpha$, a
more usual 2d N\'eel-ordered phase is stabilized, possibly with an intermediate plaquette phase.
Previous Raman experiments on this
  compound~\cite{Raman_ref} indicate that some structure in the
  spectrum can only be explained by taking into account a realistic
  model, including for instance Dzyaloshinski-Moriya (DM)
  interactions~\cite{Mazurenko2008}. However, these DM terms are too
  small to have any sizeable effects on the low-energy Raman spectra~\cite{dm} discussed 
  here and thus will be neglected hereafter.
Our goal  will be to show
that, within a simple Heisenberg model, doping non-magnetic impurities
generates new spectroscopic signatures below the 2-magnon
continuum whose features should persist in a more
realistic model.~\cite{dm} 

\begin{figure}[!ht]
\includegraphics[width=0.6\linewidth]{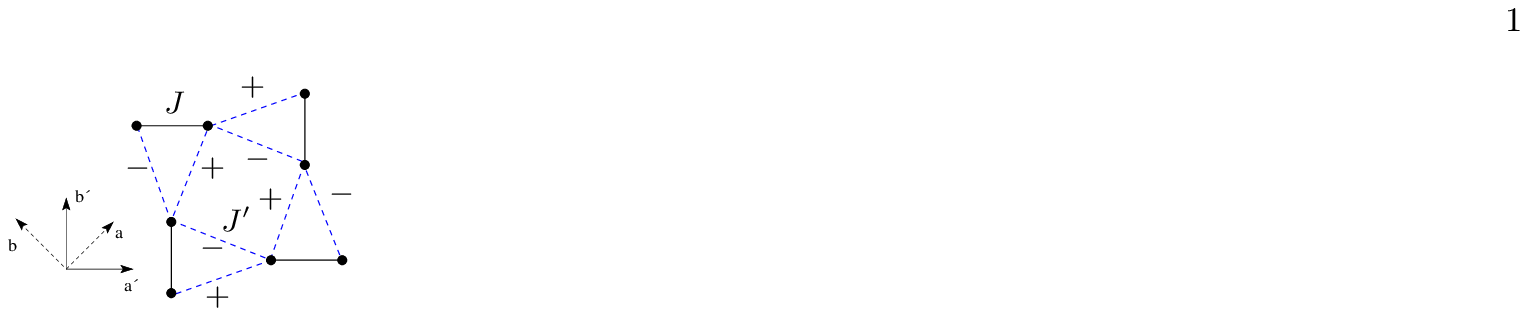}
\caption{(Color online) Shastry-Sutherland lattice. 
$\pm$ signs correspond to Raman 
coupling in the $(a'b')$ polarization case.}
\label{lattice.fig}
\end{figure}

SrCu$_2$(BO$_3$)$_2$ is a Mott insulator and its doping with mobile
carriers is predicted to lead to superconductivity~\cite{Shastry2002}.
In this paper, we shall rather consider the controlled doping of
SrCu$_2$(BO$_3$)$_2$ with non-magnetic {\it static} impurities such as
Zinc or Magnesium atoms substituted for Copper atoms on a very small
fraction of the $N$ lattice sites.  Such atoms acting as vacant sites
provide accurate local real-space probes (which can be considered as
independent for low enough impurity concentration).  As known in
strongly correlated systems, a small doping of the parent compound can
bring crucial informations about its intrinsic properties~\cite{sachdev}.
Substituting Cu with non-magnetic Mg impurities has been recently
performed~\cite{Haravifard2006,Aczel2007}.  Neutron scattering
experiments~\cite{Haravifard2006} have then shown the appearance of
new magnetic excitations into the singlet-triplet gap.  Here, we shall
focus on Raman spectroscopic techniques which, by probing $\Delta S=0$
excitations using light scattering, offer a unique way to identify the
local response to a doped impurity.  In particular, we will show that
transitions between different S=1/2 bound-states can be identified in
the Raman spectra. Such features can be interpreted within a simple
phenomenological model describing the attractive potential between a
liberated spinon and the impurity (vacant site).

\section{Theoretical framework}
\subsection{Raman scattering}
The theory of Raman light scattering is most simple when the photon energy is much smaller than the Mott gap. In such a case, it is legitimate to use the Loudon-Fleury approximation
and the Raman operator reads:
\begin{eqnarray}
{\cal R} & = & \sum_{nn} \gamma ({\bf e}_{in}\cdot {\bf d}_{ij}) ({\bf e}_{out}\cdot {\bf d}_{ij})\, {\bf S}_i \cdot {\bf S}_j \nonumber\\
&+&  \sum_{nnn}  \gamma' ({\bf e}_{in}\cdot {\bf d}_{ij}) ({\bf e}_{out}\cdot {\bf d}_{ij}) \, {\bf S}_i \cdot {\bf S}_j
\end{eqnarray}
where ${\bf e}_{in}$ and  ${\bf e}_{out}$ are the polarization vectors of the incoming and scattered
 light, and ${\bf d}_{ij}$ is the unit vector connecting two sites $i$ and $j$. This operator
only couples to zero-momentum singlet excitations. In principle, the coupling constants $\gamma$
and $\gamma'$ could depend on the exchange values~\cite{Freitas2000}, although such a dependance
is neglected in most studies.  However, 
some simplifications occur for an $(a'b')$ polarization 
(see Fig.~\ref{lattice.fig}): indeed, in this case, the geometry of the compound implies that only the
$J'$ bonds contribute to the Raman operator. Moreover, the dominant intensity occurs
when triplets can be created on dimer bonds which, in contrast, is not allowed in $(ab)$ 
polarization~\cite{Lemmens2000,Knetter2000}. In the following,  we will restrict to
 this polarization where all Raman coupling constants are equal up to a sign (given in Fig.~\ref{lattice.fig}).
 
 At zero temperature, the Raman intensity is given 
by the dynamical correlations of the Raman operator:
\begin{eqnarray}
{\cal I}_R(\omega) & = & - \frac{1}{\pi} Im \langle \Psi_0 | {\cal R} \frac{1}{\omega-{\cal H}+i\varepsilon} {\cal R} | \Psi_0\rangle \\
& = & \sum_n |\langle n | {\cal R} | \Psi_0\rangle | ^2 \, \delta(\omega-(E_n-E_0))\nonumber
\end{eqnarray}
where $|\Psi_0\rangle$ is the ground-state (GS) of the system and the sum runs over the excited
states $|n\rangle$ with energy $E_n$. Note that because of the symmetry of the Raman operator, 
only singlet states with zero momentum contribute. Moreover, in the chosen polarization, only states
which are odd with respect to reflections along $a'$ or $b'$ axis give a signal.

\subsection{Results for the pure compound}
As studied in Ref.~\onlinecite{Lemmens2000}, the Raman spectrum of the undoped
material shows 4 sharp peaks at 
1.25, 1.9, 2.3 and 2.9 times the spin gap $\Delta_{01}$. 
\begin{figure}[!ht]
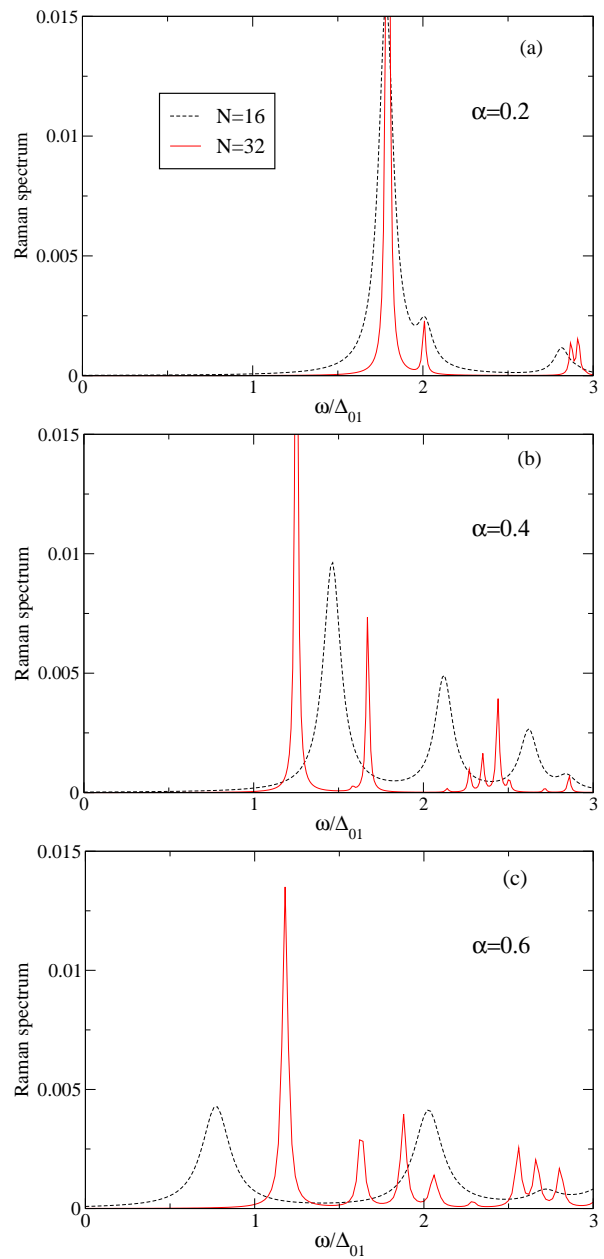

\includegraphics[width=0.9\linewidth]{fig5a}
\includegraphics[width=0.9\linewidth]{fig5b}
\includegraphics[width=0.9\linewidth]{fig5c}
\caption{(Color online) Raman spectra for pure SSL for various $\alpha$ on
$N=16$ and $32$ clusters.  
The spin gap  $\Delta_{01}$ ($\sim 0.95 J$, $0.80 J$, $0.50 J$ in (a), (b) and (c), respectively) is used as a unit of frequency. An artificial width $\varepsilon=0.01$ has been given to the delta peaks.}
\label{pure}
\end{figure}
Naively, one would expect the 
Raman spectrum to start at the 2-magnon continuum (i.e. twice the spin gap)
 since it is a singlet operator. In fact, a singlet bound state made of 2 triplets does
exist on this lattice~\cite{Knetter2000} and was identified as the low-energy state~\cite{Lemmens2000}.
The first two peaks are attributed to two-triplet bound states, while
the 2 others are interpreted in terms of three-particle excitations. 
Although the high-intensity peaks are above the 2-magnon continuum
(starting at $2\Delta_{01}$), the very small magnon dispersion
leads to a very narrow continuum so that the two higher-energy peaks are observable.

In Fig.~\ref{pure}, we present our Exact Diagonalizations spectra for various $\alpha$. As expected, for 
small $\alpha$, the finite-size effects are rather weak due to the large energy scales (i.e. short correlation lengths), so that results are almost  identical on $N=16$ and $N=32$ clusters. Clearly, spectral weight is present below the 2-magnon threshold. On the other hand, for larger $\alpha$, finite-size effects become sizeable so that a direct comparison with experimental values is difficult. However, one can still notice spectral weight well below the 2-magnon continuum, corresponding to the above mentionned singlet bound states. Note also that, for $N=32$,  the first peak is located at an energy $1.2\,\Delta_{01}$ as seen in experiments.

\subsection{Results for the doped case}
\begin{figure}[!ht]
\includegraphics[width=\linewidth,clip]{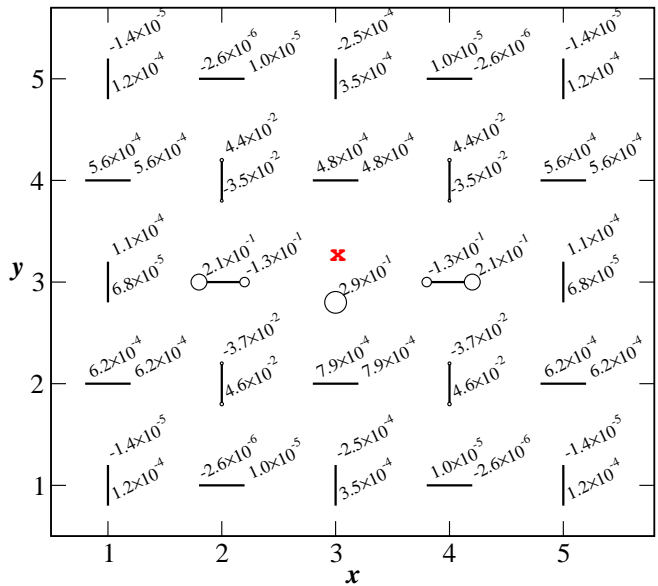}
\caption{(Color online) Magnetization pattern around the impurity (marked as x) on $N=32$ SSL for $\alpha=0.6$  ($S_z^{tot}=1/2$). Periodic boundary conditions are used. 
Since the reflection around the $b'$ axis is still a good symmetry, we observe identical values on both sides.}
\label{pattern.fig}
\end{figure}

We now turn to  the doped case for which additional low-energy states appear.
We shall assume a small enough impurity concentration so that a single impurity description
becomes legitimate. 
Using a variational approach, El Shawish and Bon\v{c}a~\cite{Shawish2006} have proposed anisotropic spin-polaronic states with a finite spatial extension around the impurity. 
Adding a single impurity (i.e. creating a vacant site by removing a spin) indeed generates a
polarization 1/2 that will distribute around the impurity. On Fig.~\ref{pattern.fig}, we show the GS magnetization pattern computed exactly on a $N=32$ cluster with one impurity.  
The local polarization oscillates from one site to the other similar to the Friedel 
oscillations reported around a localized triplet in the 1/8 plateau~\cite{Kodama2002}. 
Note however that each strong dimer has a global positive
polarization. Our results are in good agreement~\cite{compare} with the GS variational estimate of 
Ref.~\onlinecite{Shawish2006}. 
These authors have also found low-energy  states appearing below the undoped spin gap.
Since these
states are well localized, we expect that finite-size calculations can provide accurate results for these excitations as well. In particular, we expect a much better accuracy than in the pure case.


\begin{figure}[!ht]
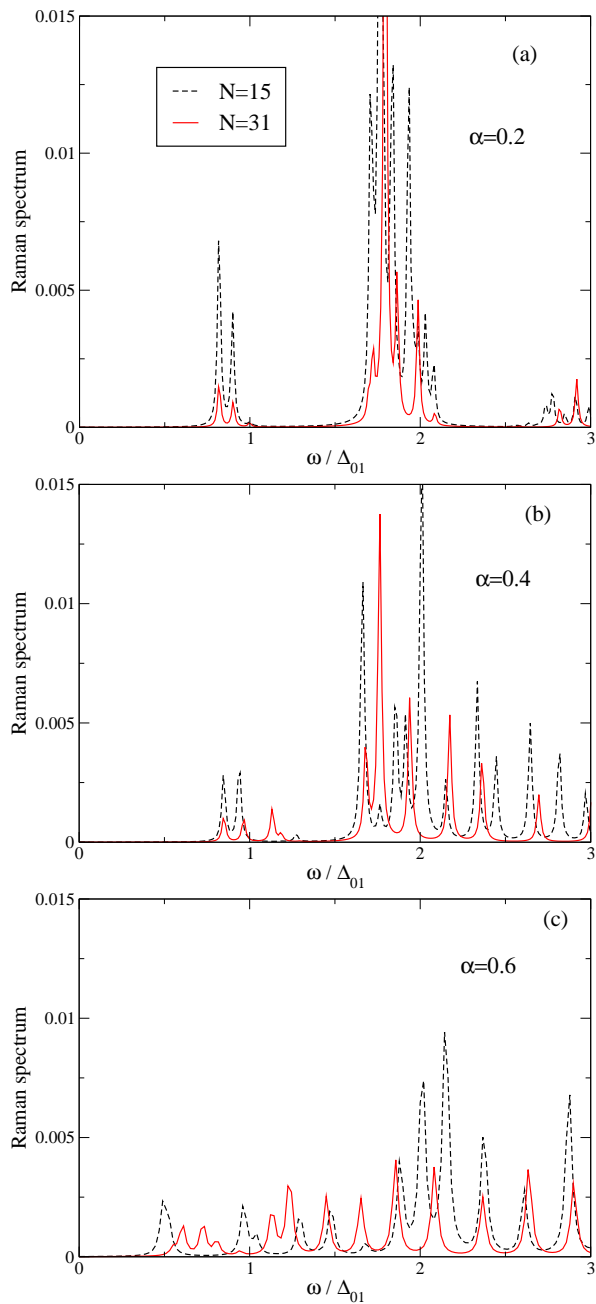

\includegraphics[width=0.9\linewidth]{fig3a}
\includegraphics[width=0.9\linewidth]{fig3b}
\includegraphics[width=0.9\linewidth]{fig3c}
\caption{(Color online) Raman spectra for SSL doped with a single non-magnetic impurity for various $\alpha$ on
$N=16$ and $32$ clusters.  
The spin gap  $\Delta_{01}$ ($\sim 0.95 J$, $0.80 J$, $0.50 J$ in (a), (b) and (c), respectively) is used as a unit of frequency. An artificial width $\varepsilon=0.01$ has been given to the delta peaks.}
\label{doped}
\end{figure}

The existence of several $S_z=1/2$ low-energy states has led to
various experimental signatures: by flipping spins, they can be
observed in neutron experiment~\cite{Haravifard2006} (which is
sensitive to $\Delta S=1$ transitions); moreover, since these states
can be connected by $\Delta S=0$ transitions, they are also expected to
be Raman active, which is the main purpose of our study.
Fig.~\ref{doped} shows the Raman spectra obtained by Exact
Diagonalizations on $N=16$ and 32 clusters with one impurity for various $\alpha$.
The spectra are plotted as a function of $\omega/\Delta_{01}$, where the spin gap of the pure sample $\Delta_{01}$ is very close
to the spin gap $\Delta_{\frac{1}{2}\frac{3}{2}}$ in the doped cluster  (both computed on the largest cluster). In the small $\alpha$ regime, all energy scales are well separated and finite-size effects are negligible so that one
can understand all features. 
For the pure system, as discussed above, there is a bound-state below $2\Delta_{01}$ 
and then another peak at $2\Delta_{01}$ corresponding to a singlet excitation made of two distant triplets. Since the
triplets have a small dispersion, there is no continuum above. 
In the presence of a single impurity, these features still represent much of the spectral weight 
but, in addition, new peaks appear below the spin gap. Namely, one can make a singlet
excitation by creating a triplet in the bulk while flipping the
polarization cloud (so that the number of such peaks scales as the number of
dimers $N_d=\frac{N}{2}-1$). Moreover, there is a possibility to form a bound-states
of these two excitations, which can even lower the energy below the spin gap (see Fig.~\ref{doped}), as found
numerically. Another feature of this additional spectral weight is
that it scales with the concentration of impurity, i.e. is reduced by
$\sim 2$ when doubling the system size.

In order to be more quantitative, we compare these bound-state
energies with the variational results of
Ref.~[\onlinecite{Shawish2006}] for $\alpha=0.62$ where the
 two lowest $S=1/2$ states have excitation
energies $0.238 J$ and $0.264 J$, while the undoped spin gap is $0.450J$. In our exact calculations on $N=32$ cluster for the same $\alpha$,
we find that the three lowest excitations have a total spin 1/2 and are
located at 0.217, 0.245 and 0.268$~J$, while the spin gap is $0.479J$ in good agreement with these
variational results, except that we have an additional low-energy state.
Concerning their Raman signatures, one has to discuss their symmetry
properties: in the presence of a single impurity, the translation
symmetry is lost and only one reflection along dimers (i.e. along $a'$
or $b'$) remain. Since the Raman operator is odd with respect to this
reflection, a simple inspection at the odd/even character of these
excited states allow to determine if they are/are not Raman active. In
 Fig.~\ref{effective.fig}(a-b), we have indicated all even and odd
$S_{tot}=1/2$ states below the first $S_{tot}=3/2$ state on $N=16$ and
$32$ clusters.  A careful comparison with the Raman spectra does
confirm that only states with $S_{tot}=1/2$ and which are odd
w.r.t. reflection give Raman peaks at low energy.

\section{Effective model for the bound-states}
\begin{figure}[!ht]
\includegraphics[width=0.9\linewidth,clip]{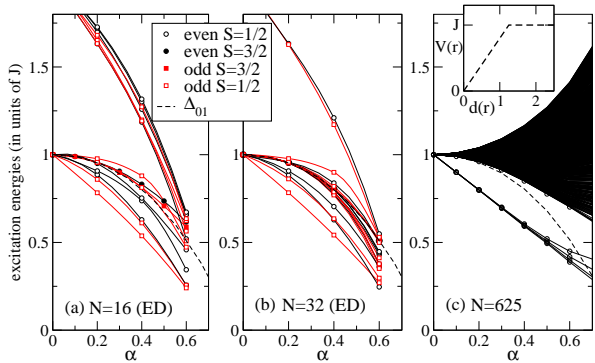}
\caption{(Color online) Low-energy S=1/2 excitations vs $\alpha$. The dashed line indicates the perturbative estimation of the spin gap (see text). (a-b) Microscopic model on $N=16$ and $32$ SSL. The states are classified according to their reflection symmetry. The two lowest S=3/2 states also shown give the onset of the continuum. (c)  Effective model (see
inset for a schematic plot of the effective potential) on a $N=625$ square lattice of dimers
for which finite-size effects are not visible. }
\label{effective.fig}
\end{figure}

Interestingly, a phenomenological description of these bound-states
can be given similarly to the case of doped quasi-1D
CuGeO$_3$~\cite{Els1998,Augier1999}.  For $J'=0$, a free $S=1/2$ (spinon) is
located next to the vacant site, on the broken dimer bond. Switching
on $J'$ allows this spinon to delocalize with a hopping term of order
$(J')^2/4J$. However, each time the spinon hops one dimer away from
the impurity, a strong bond is broken resulting in an additional
"string" energy cost $\sim (J-J')$. Therefore, the physics is similar
to a particle in a linear potential and bound-states can occur. Of
course, when the string energy exceeds the spin gap, the whole picture
breaks down and the spinon can "escape" in a flat potential by the
spontaneous creation of a spinon-antispinon pair out of the vacuum.
We have considered this effective quantum mechanical model on an
effective square lattice for one particle allowed to hop with
amplitude $\alpha^2/4$ on its neighboring sites (except between the
impurity dimer and the neighboring dimer facing the vacant site), and
with a potential energy equals to $V(r)=J\, \min((1-\alpha)d(r),1)$
where $d$ is the Manhattan distance from the impurity dimer.  This
one-particle problem can be easily solved on large clusters and its
spectrum is presented on Fig.~\ref{effective.fig}(c). For vanishing
$J'$, the low-energy S=1/2 spectrum is extensively degenerate: the
$N_d=\frac{N}{2}-1$ triplets of excitation energy $J$ located on the
remaining bonds can be combined with the impurity spin in total S=1/2
states (degenerate with their S=3/2 counter-parts not described by the
model). The next set of states which appears at energy 2J
(corresponding to 2 isolated triplets) and above are also not
described by the effective model. Switching on $J'$ lifts the
degeneracy of the first group of $N_d$ S=1/2 states resulting in a
rich spectrum well described by the effective model for which the
$N\rightarrow\infty$ limit can be taken.  Common features are observed
both in the microscopic and the effective models such as (a)
bound-states due to the short-range string-like part of the potential
and (b) a continuum of S=1/2 excitations above the spin gap (given in
Fig.~\ref{effective.fig}(a-b) by the lowest S=3/2 state).  In fact,
the estimation of the spin gap from high-order
perturbation~\cite{Miyahara1999}, $\Delta_{01} = J \left( 1-
  \alpha^2-\frac{1}{2}\alpha^3 -\frac{1}{8}\alpha^4\right)$, agrees
very well with the ED value obtained on the largest $N=32$ cluster up
to $\alpha=0.6$ and with the effective model up to $\alpha=0.3$.  For
small $\alpha$, the number of bound states below the spin gap is four:
in the effective language, it corresponds to a spinon delocalized on
one of the nearest-neighbor dimers of the impurity dimer. The
situation for larger $\alpha$ is less clear as many states go down,
possibly with stronger finite-size effects on ED data. Still, our
numerical data would be compatible with up to 12 bound-states, some of
them being very close to the spin gap. However, with our choice of
polarizations and due to the selection rule, only odd states are Raman
active, which gives 2 (resp. 6) low-energy states for $\alpha\sim 0.2$ (resp. $\alpha\sim 0.6$).

\section{Finite impurity concentration}
To finish, we quickly address the case of a finite impurity concentration 
for which a two-impurity effective interaction starts to operate. 
Since each impurity will create a spin 1/2 polarization around it, it
is natural to expect some ordering, possibly at low temperature. For
instance, antiferromagnetic (AF) ordering occurs in doped
CuGeO$_3$~\cite{renard} and has been predicted for CaV$_4$O$_9$~\cite{wessel}.  
One can imagine an
effective diluted spin 1/2 model with a very small exchange
interaction (typically, the 
overlap between two polarization clouds is exponentially small when the
impurities are quite far), and such a model is expected to order at
low-temperature.

In order to estimate the effective coupling constant between two impurities at site ${\bf r}$ and
${\bf r'}$, we start from the 
formula $J_{\rm eff}=2 (E(\uparrow \uparrow)-E(\uparrow \downarrow))$.
Now, $E(\uparrow \uparrow) = \sum_{\langle ij\rangle} J_{ij} \langle \vec S_i \cdot \vec S_j \rangle$, 
where the expectation value is calculated in the triplet state with two impurities.
Choosing the polarization along $z$, and in order to get an estimate, we can 
approximate the expectation value
by $\langle S^z_{i}\rangle \langle S^z_{j}\rangle$. 
We further approximate the polarization of a given site as
the sum of the polarizations coming from the two impurities, which would be true
to first order in perturbation, leading to 
\begin{eqnarray}
\langle S^z_{i,1}+S^z_{i,2}\rangle \langle S^z_{j,1}+S^z_{j,2}\rangle &  = &\\
 \langle S^z_{i,1}\rangle 
\langle S^z_{j,1}\rangle+\langle S^z_{i,2}\rangle \langle S^z_{j,2}\rangle &+&
\langle S^z_{i,1}\rangle \langle S^z_{j,2}\rangle + \langle S^z_{i,2}\rangle \langle S^z_{j,1}\rangle, \nonumber
\end{eqnarray}
where $\langle S^z_{i,1} \rangle$ ($\langle S^z_{i,2} \rangle$) is the average magnetization
created at site $i$ by a single impurity located at ${\bf r}$ (${\bf r'}$).
The same appromixation for the correlation functions of $E(\uparrow \downarrow)$ leads to the
same terms with the same sign for the 1-1 and 2-2 terms, and opposite signs for the 1-2 and
2-1 terms. The 1-1 and 2-2 terms will drop from the difference, leading to the formula:
\begin{equation}
J_{\rm eff}({\bf r},{\bf r'}) = 2 \sum_{\langle ij\rangle} J_{ij}  (\langle S^z_{i,1} \rangle \langle
S^z_{j,2}\rangle + \langle S^z_{i,2} \rangle \langle S^z_{j,1}\rangle)
\end{equation}
From the polarization 
shown in Fig.~\ref{pattern.fig}, the resulting effective exchange can be
obtained by considering various configurations of two impurities.
Our results (data not shown) indicate that, when both impurities are located on vertical
dimers, the effective interaction decreases very fast and is mostly
AF. Since these bonds belong to the same sublattice, the presence of
strong frustration can prevent magnetic
ordering. On the contrary, when both impurities are located on different dimer types (one vertical and one horizontal), the effective interaction is mostly ferromagnetic, thus competing with the AF and often resulting in a disordered state. Anyhow, if magnetic order occurs, it would be at a temperature
much below the effective energy scale, which is already quite
small. This argument can be generalized to an arbitrary distribution
of impurities.  In conclusion, we predict the absence of magnetic
ordering, even for quite large doping and at extremely low
temperature, which seems compatible with experiments~\cite{Aczel2007}.

\section{Conclusion}
In summary, doping a Shastry-Sutherland lattice with
non-magnetic impurities leads to novel low-energy states below the
spin gap, that could be probed by Raman spectroscopy.  We have
proposed a simple effective model to understand bound-state
formation as binding of a spinon to an impurity
site.  For a particular polarization of light, some of these
bound-states are Raman active with sizeable spectral weights: 2 of these bound-states are located well below
the continuum (and exist for any $J'/J$), while up to 4 more bound-states could be observed for  realistic parameters. 
Moreover, since the effective interactions between
impurities is frustrated, we expect no magnetic ordering down to 
extremely low temperature.

\section*{Acknowledgements}
We acknowledge B. Gaulin for very useful discussions. We thank CALMIP (Toulouse) and IDRIS (Paris) 
for allocation of cpu time. This work has been supported 
by the French Research Agency (ANR), the Swiss National Fund, by MaNEP, and the 
r\'egion Midi-Pyr\'en\'ees through its \emph{Chaire d'excellence Pierre de Fermat}. FM also
thanks LPT (Toulouse) for hospitality.


\begin{thebibliography}{10}
\bibitem{dm} In the presence of DM
    interactions, $S>0$ spin states might become Raman active. However, we are not concerned here with such excitations which
   (i) lie at higher energies than the energy range considered here and (ii) would only carry a small spectral weight.
   In addition, (iii) a perturbative DM interaction  has an even weaker effect on the Raman-active low-energy singlet-like excitations.
\end{thebibliography}
\end{document}